\documentclass[letterpaper,english,aps, prl, reprint, superscriptaddress,longbibliography]{revtex4-2}
\usepackage[T1]{fontenc}
\usepackage[latin9]{luainputenc}
\usepackage{babel}
\usepackage{natbib}
\usepackage{amsmath}
\usepackage{bbm}
\usepackage{amssymb}
\usepackage{graphicx}
\usepackage[unicode=true,pdfusetitle,
bookmarks=true,bookmarksnumbered=false,bookmarksopen=false,
 breaklinks=false,pdfborder={0 0 1},backref=false,colorlinks=false]
 {hyperref}

\makeatletter

\usepackage{hyperref}
\hypersetup{
	colorlinks = true,
	allcolors = blue
}
\usepackage{times}

\makeatother

\begin{document}
\title{Non-Abelian dynamical gauge field and topological superfluids
in optical Raman lattice}

\author{Xin-Chi Zhou}
\affiliation{International Center for Quantum Materials, School of Physics, Peking University, Beijing 100871, China}
\author{Tian-Hua Yang}
\affiliation{International Center for Quantum Materials, School of Physics, Peking University, Beijing 100871, China}
\author{Zhi-Yuan Wang}
\affiliation{International Center for Quantum Materials, School of Physics, Peking University, Beijing 100871, China}
\author{Xiong-Jun Liu}
\email{Corresponding author: xiongjunliu@pku.edu.cn}
\affiliation{International Center for Quantum Materials, School of Physics, Peking University, Beijing 100871, China}
\affiliation{Hefei National Laboratory, Hefei 230088, China}
\affiliation{International Quantum Academy, Shenzhen 518048, China}

\begin{abstract} 
We propose an experimental scheme to realize non-Abelian dynamical gauge field for ultracold fermions, which induces a novel pairing mechanism of topological superfluidity. The dynamical gauge fields arise from nontrivial interplay effect between the strong Zeeman splitting and Hubbard interaction in a two-dimensional (2D) optical Raman lattice. The spin-flip transitions are forbidden by the large Zeeman detuning, but are restored when the Zeeman splitting is compensated by Hubbard interaction. This scheme allows to generate a dynamical non-Abelian gauge field that leads to a Dirac type correlated 2D spin-orbit interaction depending on local state configurations. The topological superfluid from a novel pairing driven by 2D dynamical gauge fields is reached, with analytic and numerical results being obtained. Our work may open up a door to emulate non-Abelian dynamical gauge fields and correlated topological phases with experimental feasibility.
\end{abstract}

\maketitle

\textcolor{blue}{\em Introduction.}--Gauge theories play a fundamental role in our understanding of condensed matter~\cite{wen2007a,fradkin_2013} and elementary particles~\cite{aitchison2021}. Ultracold atoms offer versatile platforms to explore artificial gauge fields in well-controlled settings~\cite{dalibard2011,goldman2014,zhang2018}. Considerable progresses have been made in simulating static artificial gauge fields, both Abelian~\cite{jakschCreationEffectiveMagnetic2003,lin2009,aidelsburger2011,aidelsburger2013,miyake2013,kennedy2015,liu2009,lin2011,wang2012,cheuk2012,struck2013,jotzu2014,kolkowitz2017,liu2013,song2018} and non-Abelian~\cite{osterloh2005a,ruseckas2005a,huang2016,liu2014,wu2016,wang2018,sun2018,song2019,liang2023,lu2020,wang2021},
in the past years.
Fundamentally, the gauge fields are dynamical rather than static, and rely on the configuration of local matter fields which back act on the gauge fields~\cite{banuls2020}.
Recent experiments have successfully realized in quantum simulators dynamical gauge fields with or without local gauge symmetries, of which the former render the lattice gauge theory models~\cite{martinez2016,schweizer2019,mil2020b,yang2020,zhou2022,banuls2020}, and the latter are typically realized with dynamical dependence on local density of matter particles~\cite{clark2018,gorg2019,kroeze2019,lienhard2020,yao2022,rosa-medina2022,frolian2022}, 
known as density-dependent gauge fields. 

Thus far the dynamical gauge fields emulated in experiments have been restricted in Abelian type, and the extension to non-Abelian case remains to be experimentally challenging but is of great interests. 
Even the Abelian density-dependent dynamical gauge fields lead to various novel many-body phenomena, including dynamical one-dimensional (1D) spin-orbit effects~\cite{xu2018a,kroeze2019,xu2021}, anyon Hubbard models~\cite{keilmann2011,greschner2015,strater2016,kwan2023}, chiral solitons~\cite{edmonds2013}, and novel quantum phase transitions~\cite{greschner2014,gonzalez-cuadra2018,gonzalez-cuadra2020,colella2022}. 
These studies bring an urgent question to answer: How to realize non-Abelian density-dependent dynamical gauge field in an experimental model, and what novel many-body phases may be feasibly achieved from such dynamical gauge field? We address this important issue by combining strong interaction with optical Raman lattice~\cite{liu2013,liu2014,pan2015,liu2016,wang2018,zhang2018,lu2020,zhang2016,zhou2017,liu2019,ziegler2022b,cai2022,zhang2022,cao2022,li2023}, a platform that has widely advanced the experimental progresses in exploring topological quantum physics with static high-dimensional spin-orbit couplings (SOCs)~\cite{wu2016,sun2018,sun2018a,song2019,yi2019,wang2021,liang2023,zhang2023a}. 

We propose a feasible scheme to realize 2D non-Abelian dynamical gauge fields, and uncover a novel mechanism for topological superfluidity (TSF) which is highly-sought-after but still extremely challenging for experiment~\cite{tewari2007,zhang2008,sato2009,zhu2011,zhou2011,alicea2012,gong2012b,qu2013,zhang2013,liu2014,cao2014,liu2014a,wu2016a,hu2018,poon2018,jia2019,zeng2019,chen2022a,huang2022}. Our realization is based on a nontrivial compensation effect shown here between strong Hubbard interaction and large Zeeman detuning in 2D optical Raman lattice. The single-particle spin-flip transitions are suppressed by the large Zeeman splitting, but the two-particle processes between doublon $\lvert\uparrow\downarrow\rangle_{\vec i}$ and neighboring spin-down states $\lvert\downarrow\rangle_{\vec i}\lvert\downarrow\rangle_{\vec i\pm1}$ can be assisted by strong Hubbard interaction which compensates the Zeeman detuning. This leads to correlated non-Abelian SOC terms and a new pairing mechanism for TSF which is driven by real-space dynamical gauge field rather than by the nesting of single-particle Fermi surfaces. The prediction is of high feasibility. First, the non-Abelian dynamical gauge field and associated TSF phase emerge generally whenever the Hubbard interaction and Zeeman detuning compensate, regardless of how strong they are. Secondly, the large Zeeman detuning projects out all other competing orders, leaving only a pure TSF phase in the relevant fillings which can be easily achieved in experiment. Finally, the magnetic noise that induces random single-particle spin-flip couplings in the conventional SOC systems can be significantly suppressed by the large Zeeman splitting in the present system. This makes our current scheme be robust to related environment fluctuations. 

\begin{figure}[t]
\includegraphics{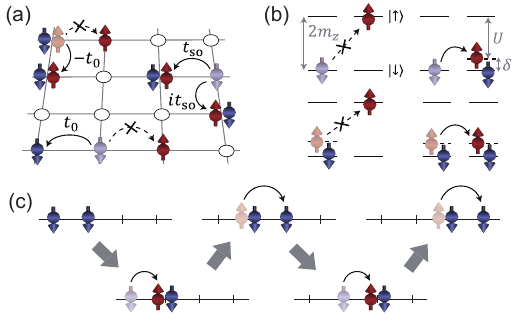}\caption{\label{fig:DDSOC}Non-abelian dynamical gauge field in an optical Raman lattice. (a) The spin-conserved and spin-flip hopping processes become density-dependent in the strongly correlated regime $m_{z},U\gg t_{0},t_{\mathrm{so}}$
with $U\approx2m_{z}$. Hopping processes that create singly occupied
spin up states $\lvert\uparrow\rangle$ are prohibited. Spin-up particles can only appear in pairs with spin-down particles on the same site. (b) Mechanism of density-dependent SOC.
The spin-flip hopping is suppressed by the large Zeeman detuning. The resonant attractive interaction compensates this energy offset and drives density-dependent spin-flip hopping (upper panel).
Spin-conserved hopping of spin-up particles only occurs in the presence of a spin-down particle on the same site (lower panel).
(c) Schematic of an example of pair hopping via the correlated SOC.}
\end{figure}

\textcolor{blue}{\em Scheme.}--We consider a 2D SOC model in the presence of a large Zeeman field and a strong attractive $s$-wave Hubbard interaction, which is written as
\begin{equation}
H=H_{\mathrm{SOC}}+H_{\mathrm{Zeeman}}+H_{\text{\ensuremath{\mathrm{int}}}}-\mu\sum_{\vec{i}}(\hat{n}_{\vec{i}\uparrow}+\hat{n}_{\vec{i}\downarrow}),
\end{equation}
with $\mu$ the chemical potential and $H_{\text{\ensuremath{\mathrm{SOC}}}}$ given by
\begin{equation*}
H_{\mathrm{SOC}}=\sum_{\langle\vec{i},\vec{j}\rangle,s,s^\prime}
\mathbf{t}\big(\vec{i}-\vec{j}\big)\cdot\boldsymbol{\sigma}_{ss^\prime}c_{\vec{i}s}^{\dagger}c_{\vec{j}s^\prime},
\end{equation*}
where $c_{\vec{i}s}^{\dagger}$ ($c_{\vec{i}s}$) creates (annihilates) a fermion with spin $s=\{\uparrow,\downarrow\}$
on site $\vec{i}=(i_{x},i_{y})$. $\boldsymbol{\sigma}=(\sigma_x,\sigma_y,\sigma_z)$ are the Pauli matrices. The hopping coefficient takes 2D form $\mathbf{t}(\pm \hat{x})=(0, \mp i t_{\mathrm{so}}, -t_0)$, $\mathbf{t}(\pm \hat{y})=(\mp i t_{\mathrm{so}}, 0, -t_0)$, and zero otherwise, with $t_{0}$ ($t_{\mathrm{so}}$)
denoting the spin-conserved (spin-flip) hopping coefficient. The Zeeman term is given by
\begin{equation*}
    H_{\mathrm{Zeeman}}=m_{z}\sum_{\vec{i}}(\hat{n}_{\vec{i}\uparrow}-\hat{n}_{\vec{i}\downarrow}),
\end{equation*}
with $\hat{n}_{i\sigma}=c_{\vec{i}\sigma}^{\dagger}c_{\vec{i}\sigma}$ and $m_z$ the strength of Zeeman splitting, and the attractive Hubbard interaction, which can be tuned by Feshbach resonance~\cite{bloch2008}, is given by
\begin{equation*}
    H_{\mathrm{int}}=-U\sum_{\vec{i}}\hat{n}_{\vec{i}\uparrow}\hat{n}_{\vec{i}\downarrow},
\end{equation*}
with $U>0$ being the strength of interaction. When $U=0$ and $|m_{z}|<4t_{0}$, the Hamiltonian $H$
describes a quantum anomalous Hall (QAH) model which has been studied thoroughly in both theory and
experiment in the optical Raman lattice~\cite{liu2014,wu2016,wang2018,sun2018,sun2018a}.

In this study, we focus on the strongly correlated regime with $m_{z},U\gg t_{0},t_{\mathrm{so}}$ while $U\sim2m_{z}$, in which regime $H_\mathrm{SOC}$ shall become density-dependent and give rise to a non-Abelian dynamical gauge field as depicted in Fig.~\ref{fig:DDSOC}(a).
We first consider a minimal two-particle case as shown in Fig.~\ref{fig:DDSOC}(b). When $U=0$, the single-particle spin-flip transitions are suppressed by large Zeeman detuning $2m_z$, and the ground state is fully polarized to $\lvert \downarrow\rangle$. Introducing attractive interactions with $U\approx2m_z$ compensates the detuning between $\lvert\uparrow\downarrow\rangle_{\vec i}$ and $\lvert\downarrow\rangle_{\vec{i}} \lvert\downarrow\rangle_{\vec i\pm1}$ and drives density-dependent spin-flip process.
Thus the spin-up fermions appear only in doublons through the dynamical SOC on two neighboring spin-down atoms, rendering a key feature of our realization.
For the 2D lattice, we reach the effective Hamiltonian to the lowest order $H_{\mathrm{eff}} = P_{\uparrow}(H_{\mathrm{SOC}}+\tilde{H}_{\mathrm{Zeeman}})P_{\uparrow}$~\cite{Supplement}, where $P_{\uparrow}=\prod_{\vec{i}}[1-\hat{n}_{\vec{i}\uparrow}(1-\hat{n}_{\vec{i}\downarrow})]$ is the projection that excludes singly occupied spin-up states and $\tilde{H}_{\mathrm{Zeeman}}=\frac{\delta}{2}\sum_{\vec{i}}(\hat{n}_{\vec{i}\uparrow}-\hat{n}_{\vec{i}\downarrow})$ characterizes
an effective detuning between doublon $\lvert \uparrow\downarrow,0\rangle$ and paired spin-down states $\lvert\downarrow,\downarrow\rangle$, with $\delta/2=2m_{z}-U$.
Equivalently, the Hamiltonian can be recasted into a density-dependence of the gauge fields
\begin{equation}\label{Heff}
  H_{\mathrm{eff}} = \sum_{\langle\vec{i},\vec{j}\rangle,s,s^\prime}c_{\vec{i}s}^{\dagger}
  \begin{bmatrix}\mathcal{\mathcal{A}}_{\vec{i},\vec{j}}^{11}(\hat{n}) & \mathcal{\mathcal{A}}_{\vec{i},\vec{j}}^{12}(\hat{\mathbf{r}},\hat{n})\\
    \mathcal{\mathcal{A}}_{\vec{i},\vec{j}}^{21}(\hat{\mathbf{r}},\hat{n}) & \mathcal{\mathcal{A}}_{\vec{i},\vec{j}}^{22}(\hat{n})
  \end{bmatrix}
  c_{\vec{j}s^\prime}+\tilde{H}_{\mathrm{Zeeman}},
  \end{equation}
where $\mathcal{A}^{11}_{\vec{i},\vec{j} }(\hat{n})=-\hat{n}_{\vec{i},\downarrow}\hat{n}_{\vec{j}\downarrow}t_0$ and $\mathcal{A}^{22}_{\vec{i},\vec{j}}(\hat{n})=(1-\hat{n}_{\vec{i}\uparrow})(1-\hat{n}_{\vec{j}\uparrow})t_0$ are gauge field operator for spin-conserved processes, and $\mathcal{A}^{12}_{\vec{i},\vec{j} }(\hat{\mathbf{r}},\hat{n})=a(\mathbf{\hat{r}})\hat{n}_{\vec{j} \downarrow}(1-\hat{n}_{\vec{i} \uparrow})t_{\mathrm{so}}$, $\mathcal{A}^{21}_{\vec{i},\vec{j} }(\hat{\mathbf{r}},\hat{n})=-a(\hat{\mathbf{r}})^{*}\hat{n}_{\vec{i} \downarrow}(1-\hat{n}_{\vec{j} \uparrow})t_{\mathrm{so}}$ for spin-flip processes. Here $\hat{\mathbf{r}}=\{\pm\hat{x},\pm\hat{y}
\}$ is the direction unit vector, $a(\pm\hat x)=\mp 1$ and $a(\pm\hat y)=\mp i$. The realized 2D correlated SOC associated with real-space dynamical gauge field enables an effective pair hopping channel of doublon and paired spin-down states as illustrated in Fig.~\ref{fig:DDSOC}(c). This result inspires us to explore a new pairing mechanism for TSF resulting from the non-Abelian dynamical gauge field.

\textcolor{blue}{\em Two-body problem.}--
To elucidate the non-trivial pairing in the dynamical gauge field, we first identify the two-body bound state by solving the two-body problem~\cite{vyasanakere2011,hu2011,yu2011,dong2013}. We adopt the two-body trial wave function as a mixture of  $s$- and $p$-wave pairs taking the form
\begin{equation}\label{eq:TBTrial}
\left|\Psi_{\vec{q}}\right\rangle =\sum_{\vec{k}}\left[\psi_{s}c_{\frac{\vec{q}}{2}+\vec{k}\uparrow}^{\dagger}c_{\frac{\vec{q}}{2}-\vec{k}\downarrow}^{\dagger}+\psi_{p}(\vec{k})c_{\frac{\vec{q}}{2}+\vec{k}\downarrow}^{\dagger}c_{\frac{\vec{q}}{2}-\vec{k}\downarrow}^{\dagger}\right]|0\rangle,
\end{equation}
where $\vec{q}$ is the center-of-mass momentum.
Considering the projection operator $P_\uparrow$ that eliminates the single-occupation of spin-up particles, the Schr{\"o}dinger equation is written as $H_{\mathrm{eff}}P_\uparrow\lvert \Psi_{\vec{q}}\rangle=E\lvert\Psi_{\vec{q}}\rangle$ 
yields~\cite{Supplement},
\begin{equation}
  E=f(E):=-8t_{\mathrm{so}}^{2}\iint_{K}\frac{\mathrm{d}^{2}\vec{k}}{4\pi^{2}}\frac{(\sin^{2}k_{x}+\sin^{2}k_{y})\cos^{2}\frac{q_{x}}{2}}{2E_{0}(\vec{k})-E},\label{eq:Self-Con-Cond}
\end{equation}
and
\begin{equation}
\frac{\psi_{p}(\vec{k})}{\psi_{s}}=-2it_{\mathrm{so}}\frac{(\sin k_{x}-i\sin k_{y})\cos \frac{q_{x}}{2}}{2E_{0}(\vec{k})-E}. \label{eq:PSRatio}
\end{equation}
Here $2E_{0}(\vec{k})=4t_{0}(\cos k_{x}+\cos k_{y})\cos (q_{y}/2)-\delta$, the integration domain $K$ is $\{\vec{k}|E_{0}(\vec{k})>E_F\}$, with the Fermi level $E_F<0$ denoting the less-than-half filling case. A bound state with $E<2E_F$ always exists as shown by the asymptotic behavior of Eq.~\eqref{eq:Self-Con-Cond}. When $E\rightarrow -\infty$, $f(E)\rightarrow 0$ and when $E= 2E_F+0^{-}$, $f(E)$ diverges logarithmically to $-\infty$. Hence, as $E$ varies from $-\infty$ to $2E_F+0^{-}$, $E-f(E)$ changes from $-\infty$ to $+\infty$, implying a bound state solution $E<2E_F$ satisfying $E=f(E)$. Moreover, $\psi_{p}(\vec{k})$ has an approximate $k_x-ik_y$ profile as evident from Eq.~\eqref{eq:PSRatio}, which agrees with the two-body ansatz in Eq.~\eqref{eq:TBTrial}. These results suggest the formation of a bound state as a mixture of $s$- and $p$-wave pairs.

\begin{figure}[t]
\includegraphics{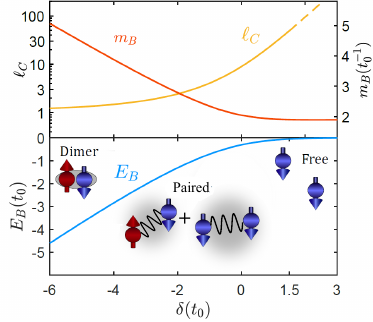}

\caption{\label{fig:TBPhy}Self-consistent two-body calculations 
for $\vec{q}=0$. The binding energy $E_{B}$ , pair size $\ell_{C}$
and effective mass $m_{B}$ are shown as functions of the effective Zeeman splitting $\delta$. The bound state exhibits three distinct regimes.
At large negative $\delta$, the pair is $s$-wave dominated, localized and immobile, forming a molecule. At large positive $\delta$, $E_B$ tends to zero, the pair becomes large and weak, tending to dissolve into two free spin-down particles.
Around $\delta\approx0$, a mobile bound state with both $s$- and $p$-wave components emerges. 
}
\end{figure}

We further reveal the correlated nature of mobile $s$- and $p$-wave pairs by investigating the binding energy $E_{B}=E-2E_{F}$, the pair size $\ell_{C}$ and the effective mass $m_{B}$ of the bound state at $\vec{q}=0$. Fig.~\ref{fig:TBPhy} shows that the binding strength $|E_B|$ decreases with increasing $\delta$. Moreover, the relative amplitude $|\psi_{p}(k)/\psi_{s}|$, which reflects the relative contribution of $s$- and $p$-wave components, increases as the $|E_B|$ weakens, as shown in Eq.~\eqref{eq:PSRatio}. Thus, the bound state evolves from an $s$- to $p$-wave dominated state as $\delta$ increases. Remarkably, a mobile paired state is always obtained in the near resonance regime $\delta\sim0$, where both $s$- and $p$-wave components are present. In contrast, for $\delta/t_0\ll-1$, a heavy effective mass characterizes the $s$-wave dominated state, giving a localized and immobile molecule; whereas for $\delta/t_0\gg1$ a low $E_B$ with large pair size indicates a dissociation of the bound state for the $p$-wave limit regime. This suggests the existence of a novel form of superfluidity enabled by the interplay of $s$- and $p$-wave pairings.

\begin{figure}[t]
\includegraphics{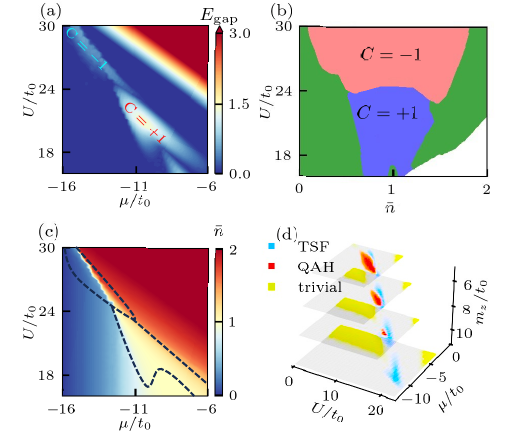}
\caption{\label{fig:MF}Phase diagram from mean-field calculations with $t_0=t_{\mathrm{so}}$. (a) Bulk
gap $E_{g}$ versus $\mu$ and $U$ with $m_{z}=12t_{0}$. (b) The phase diagram obtained by replacing the $\mu$ by self-consistent filling number $\bar n$. The green regime is either gapless Fermi gases or trivial insulator. (c) Self-consistent filling number $\bar n$ with
$m_{z}=12t_{0}$, dashed area corresponds to the TSF phases. (d) Phase diagrams versus $\mu$ and $U$ under various Zeeman
detunings $m_{z}=5t_{0}$, $7t_{0}$, $9t_{0}$ and $11t_{0}$. The TSF,
QAH and trivial insulator phase are marked as blue, red and yellow,
respectively. The QAH region shrinks as $m_z$ increases and eventually disappears as discussed in the text. }
\end{figure}

\begin{figure*}[th]
  \includegraphics{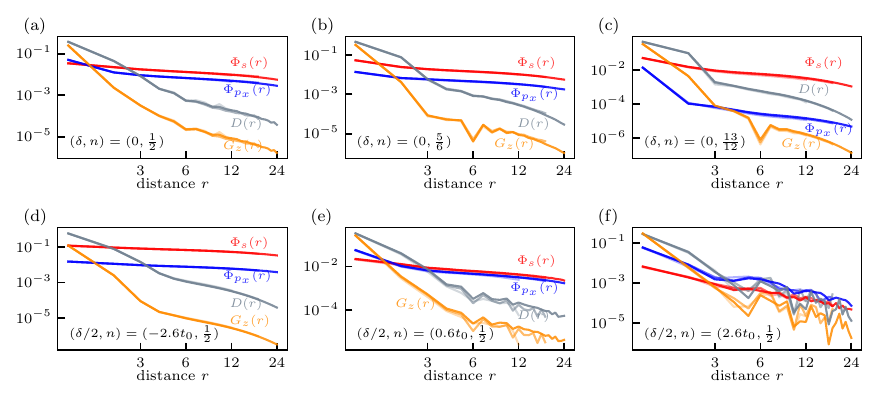}
  \caption{\label{fig:DMRG} Evidence for superfluidity. Correlation functions for different
  fillings (a-c) and detunings (d-f). Double-logarithmic plot of
  pair-pair $\Phi_{s/p}(r)$, density-density $D(r)$ and spin-spin correlation
  function $G_z(r)$ with $t_{0}=t_{\mathrm{so}}$. (a)-(c) correspond to $n=\frac{1}{2},\frac{5}{6},\frac{13}{12}$, respectively, with $\delta=0$. $\Phi_{s/p}(r)$ exhibit the slowest algebraic decay, indicating superfluidity. (d)-(f) correspond
  to $\delta=-2.6t_{0},0.6t_{0},2.6t_{0}$, respectively, at $n=\frac{1}{2}$. Negative $\delta $ enhances pair correlations and positive $\delta $ weakens them. In our DMRG simulation, we keep up to 1200 states to obtain accurate results with truncation error $\epsilon\leq5\times10^{-6}$ on cylinders with $L_{y}=4$. Line colors from light to dark represent $L_x=24$, $36$ and $48$.}
\end{figure*}

\textcolor{blue}{\em Phase diagram.}--We next study the new pairing mechanism and self-consistent phase diagram. 
The essential order
parameters are $s$ ($p$)-wave superfluid order parameter $\Delta_{s}$ ($\Delta_{p}$) and average density of each spin species $\bar n_{\sigma}$, which are given by $ \Delta_{s}=\frac{1}{N_{0}}\sum_{\vec{i}}\langle c_{\vec{i}\uparrow}^{\dagger}c_{\vec{i}\downarrow}^{\dagger}\rangle$, $\Delta_{p}=\frac{1}{N_{0}}\sum_{\vec{i}}\langle c_{\vec{i}\downarrow}^{\dagger}c_{\vec{i}+\vec{e}\downarrow}^{\dagger}\rangle$ and $\bar n_{\sigma}=\frac{1}{N_{0}}\sum_{i}\langle\hat{n}_{i\sigma}\rangle$, respectively. Then the matter-gauge coupling term can be approximated as
\begin{align}
  t_{\mathrm{so}}\text{\ensuremath{\hat{n}}}_{\vec{i}\downarrow}c_{\vec{i}\uparrow}^{\dagger}c_{\vec{i}+\hat{\mathbf{r}}\downarrow}(1-\hat{n}_{\vec{i}+\hat{\mathbf{r}}\uparrow})\approx t_{\mathrm{so}}\bar{n}_{\downarrow}(1-\bar{n}_{\uparrow})c_{\vec{i}\uparrow}^{\dagger}c_{\vec{i}+\hat{\mathbf{r}}\downarrow}	\nonumber\\
  -t_{\mathrm{so}}\Delta_{s}(1-\bar{n}_{\uparrow})c_{\vec{i}+\hat{\mathbf{r}}\downarrow}c_{\vec{i}\downarrow}-t_{\mathrm{so}}\Delta_{p}(1-\bar{n}_{\uparrow})c_{\vec{i}\uparrow}^{\dagger}c_{\vec{i}\downarrow}^{\dagger},	
\end{align} where the first term is renormalized spin-flip hopping, and
the last two terms describe coupling of doublon and paired spin-down state as depicted in Fig.~\ref{fig:DDSOC}(c). We calculate the bulk gap $E_{\mathrm{gap}}$ and the mean-field phase diagram as a function of $U$ and $\mu$ for $m_z=12t_0$ and $t_0=t_{\mathrm{so}}$, and obtained the TSF phases with both Chern number $\mathrm{C}=1$ and $\mathrm{C}=-1$ for a broad range of $U$ and $\mu$ [Fig.~\ref{fig:MF}(a)]. The corresponding self-consistent $\Delta_s$ and $\Delta_p$ are given in the Supplementary Materials~\cite{Supplement}. Outside the TSF region, the ground state manifests as either a gapless Fermi gas
or a trivial band insulator. 
We also present the phase diagram in terms of average particle number $\bar n$, which is truly relevant for experiments. As shown in Fig.~\ref{fig:MF}(b,c), the $\mathrm{C}=-1$ TSF phase actually covers a wide range of particle filling with fixed $U$, unlike the narrow parameter range of $\mu$. In particular, starting from the large $\delta$ regime which corresponds to fully polarized band insulator in the $\lvert\downarrow\rangle$ state, the TSF phase with $\bar n\approx1$ can be feasibly prepared in experiment by varying $\delta$ to the proper small magnitude. 

The above results show also a crucial feature that in the large $m_z$ regime the pure TSF is obtained without competing orders at relevant fillings, which is further systematically verified from the phase diagrams in Fig.~\ref{fig:MF}(d) by varying $m_z$.  
We find that the QAH phase exists for small and moderate $m_z$, while disappears for large $m_z$, leaving only the TSF phase. This important result is a consequence of different underlying mechanisms of QAH and TSF in the different regimes. The QAH results from single-particle SOC at half-filling, which is restored at small and moderate $m_z$. Then the QAH gap may open if the Hubbard interaction corrects the Zeeman detuning as $\tilde{m}_z=m_z - U(n_{\uparrow}-n_{\downarrow})/2$ to small enough values~\cite{Supplement}. 
However, at large $m_z$ regime, the single-particle SOC is projected out, while the correlated SOC exists, so the QAH phase disappearS in general. In sharp contrast, the pairing order results from the 2D correlated SOC in the large $m_z$ regime, 
and the TSF is always obtained at relevant fillings when $m_z$ and $U$ satisfy the near resonance condition. 
This novel mechanism ensures a high feasibility of realizing TSF at large Zeeman detuning regime.

The topological phase transition with Chern number changing by two can be analytically examined near the resonance point $U=2m_{z}$ at the average filling $\bar n=1$. 
Near the phase boundary, the average particle number $\bar n$ approaches one. The effective chemical potential
$\mu_{\mathrm{eff}}=\mu+U\bar n/2$ approaches zero, resulting in a linear dependence between $\mu$ and $U$ that
$\mu\approx-U/2$. At the two symmetric points $\vec{k}=(0,\pi)$ and $\vec{k}=(\pi,0)$ the bulk gap closes for $(m_{z}-U/2)(m_{z}-U/2+2U\bar n_{\uparrow})=0$. Then the change of 2 in the Chern number of TSF phase originates from the fact that tuning $U$ across $2m_{z}$ reverses the mass signs at both $\vec{k}=(0,\pi)$ and $\vec{k}=(\pi,0)$ [\ref{fig:MF}(b,c)]. In addition, the bulk gap also closes at $\vec{k}=(0,0)$ and $\vec{k}=(\pi,\pi)$ for $(m_{z}\mp 4t_0 -U/2)(m_{z} \mp 4t_0-U/2+2U\bar n_{\uparrow})=0$. These results further provide a qualitative picture for the TSF phases away from the average half-filling regime. The feasibility of preparing the system at half-filling, as pointed out above, also facilitates the realization of 
TSF around $\bar n=1$ with different Chern numbers.

\textcolor{blue}{\em DMRG analysis.}--
We verify the pairing orders obtained in mean-field theory as dominant correlations by density matrix renormalization group (DMRG) on narrow cylinder. We consider periodic (open) boundary condition in the $y$ $(x)$ direction of the cylinder, with width $L_{y}$ and length $L_{x}$, respectively. The ground state properties with dynamical gauge field in Eq.~(\ref{Heff}) are examined by comparing the decay behavior of different correlation functions using DMRG~\cite{white1992,schollwock2011}. We define the pair-pair correlation function as $\Phi_{\eta}(r) =\sum_{y=1}^{L_{y}}\langle \hat \Delta_{\eta}^{\dagger}(x_{0},y)\hat \Delta_{\eta}(x_{0}+r,y)\rangle /L_y $, with $\eta=\{s,p\}$ and $\hat \Delta_{s}^{\dagger}(i,j)=c_{i,j\uparrow}^{\dagger}c_{i,j\downarrow}^{\dagger}$, $\hat \Delta_p^{\dagger} (i,j)=c_{i,j\downarrow}^{\dagger}c_{i+\hat{x},j\downarrow}^{\dagger}$, the density-density correlation function as $D(r)=\sum_{y=1}^{L_{y}}(\langle\hat{n}_{x_{0},y}\hat{n}_{x_{0}+r,y}\rangle-\langle\hat{n}_{x_{0},y}\rangle\langle\hat{n}_{x_{0}+r,y}\rangle)/L_{y}$ and the spin-spin correlation function as $G_z(r)=\sum_{y=1}^{L_{y}}(\langle S^z_{x_{0},y}S^z_{x_{0}+r,y} \rangle-\langle S^z_{x_{0},y} \rangle\langle S^z_{x_{0}+r,y} \rangle)/L_{y}$. Here $S^z_{x,y}$ is the magnetization operator on site $(x,y)$, $(x_{0},y)$ is the reference site taken as $x_0=L_x/4$ and $r$ is the distance along $x$-direction. Fig.~\ref{fig:DMRG} shows the correlation functions for cylinders with $L_{y}=4$ at different fillings $n$ and effective detuning $\delta/2=2m_z-U$. Close to resonant regime $\delta=0$, pair-pair correlations decay much more slowly than $D(r)$ and $G_z(r)$ at various $n$ [Fig.~\ref{fig:DMRG}(a)-(c)], indicating the expected dominance of the pairing orders. We note that a stronger $\Phi_s(r)$ than $\Phi_p(r)$ does not mean the ground state to be topologically trivial; instead it is still in the TSF phase, which is driven by a combination of $s$-wave pairing and SOC at the proper fillings. We then fix $n=1/2$ and investigate the behavior of pairing order away from resonant condition. When $\delta<0$, the amplitude of $\Delta_s(r)$ increases [Fig.~\ref{fig:DMRG}(d)], showing the enhancement of $s$-wave pairing by attractive interaction. When $\delta>0$, pairing correlations also persist up to moderate $\delta$ [Fig.~\ref{fig:DMRG}(e)]. This is consistent with the mean-field results that $U>2m_z$ covers a broader range of $\bar n$ for TSF phase as shown in Fig.~\ref{fig:MF}(c). When $\delta$ is too large, all the correlations decay similarly in the long distance, manifesting that the pairing order also disappears with the overlarge Zeeman detuning [Fig.~\ref{fig:DMRG}(f)]. These results are consistent with the mean-field calculation. 

\textcolor{blue}{\em Conclusion.}--We have proposed a scheme to generate non-Abelian dynamical gauge field for ultracold fermions and unveiled a novel pairing mechanism for TSF phases. The single particle SOC is prohibited by the large Zeeman detuning but a correlated SOC can be assisted by the near resonant attractive Hubbard interaction. This results in a density dependent dynamical SOC with effective pair hopping processes. A pure TSF phase without other competing phases can be achieved feasibly at the relevant fillings. Furthermore, we demonstrate that the TSF phase is robustly obtained under the variation of large Zeeman splitting, as long as the near resonant compensation by interaction is met. Our scheme also avoids the detrimental effects of environmental noise fluctuations on the single-particle spin-flip transitions, can greatly enhance the capability of engineering synthetic dynamical non-Abelian gauge fields in quantum simulators, and may open an avenue to realize exotic correlated topological phases, including the 
highly-sought-after TSF phase with experimental feasibility.

\textcolor{blue}{\em Acknowledgments.}--We are grateful for valuable discussions with Sen Niu, Lin Zhang, Ke Wang and Ting-Fung Jeffrey Poon. The DMRG algorithm is implemented based on the ITensor library~\cite{fishmanITensorSoftwareLibrary2022}. This work was supported by National Key Research and Development Program of China (2021YFA1400900), the National Natural Science Foundation of China (Grants No. 11825401 and No. 12261160368), the Innovation Program for Quantum Science and Technology (Grant No. 2021ZD0302000), and the Strategic Priority Research Program of the Chinese Academy of Science (Grant No. XDB28000000).

%
  
\renewcommand{\thesection}{S-\arabic{section}}
\setcounter{section}{0}  
\renewcommand{\theequation}{S\arabic{equation}}
\setcounter{equation}{0}  
\renewcommand{\thefigure}{S\arabic{figure}}
\setcounter{figure}{0}  
\renewcommand{\thetable}{S\Roman{table}}
\setcounter{table}{0}  
\onecolumngrid \flushbottom 

\newpage
\begin{center}\large \textbf{Supplementary Material} \end{center}

\section{Effective Hamiltonian}
\label{sec:Hameff}
In this section, we derive the effective Hamiltonian in the main text. The original tight-binding Hamiltonian is given by
\begin{eqnarray}
  H&=& H_\mathrm{SOC} + H_\mathrm{Zeeman} +
  H_\mathrm{int}
  \nonumber\\
  & = & - t_0 \sum_{<\vec{i},\vec{j}>} (
  c^{\dag}_{\vec{i} \uparrow} c_{\vec{j} \uparrow} -
  c^{\dag}_{\vec{i} \downarrow} c_{\vec{j} \downarrow} ) +
  \sum_{<\vec{i},\vec{j}>} (
  t_\mathrm{so}^{\vec{i} \vec{j}} c_{\vec{i}
  \uparrow}^{\dag} c_{\vec{j} \downarrow} +\mathrm{h.c.} )
  \nonumber\\
  &  & + m_z \sum_{\vec{i}} ( \hat{n}_{\vec{i} \uparrow} -
  \hat{n}_{\vec{i} \downarrow} ) - U \sum_{\vec{i}}
  \hat{n}_{\vec{i} \uparrow} \hat{n}_{\vec{i} \downarrow}.
\end{eqnarray}
As mentioned in the main text, we focus on the strongly correlated
regime $U \approx 2 m_z \gg t_0$($t_\mathrm{so}$), in which singly
occupied spin-up state $\lvert \uparrow \rangle$ takes highest on-site energy $m_z$. So
we can classify the hopping events into three types, processes that leads to
(i) an increase, (ii) a decrease and (iii) conservation of the number of
$\lvert \uparrow \rangle$. Thereby we decompose the $H_\mathrm{SOC}$ into
three terms
\begin{equation}
  H_\mathrm{SOC} = H_\mathrm{SOC}^+ + H_\mathrm{SOC}^- +
  H_\mathrm{SOC}^0,
\end{equation}
with $H_t^+$ includes all the hopping processes that {\em increase} the
number of $\lvert \uparrow \rangle$ are given by
\begin{eqnarray}
  H_\mathrm{SOC}^+ & = & - t_0 \sum_{<\vec{i},\vec{j}>} \sum_{\sigma} \left[ 2 \sigma ( 1 - \hat{n}_{\vec{i}
  , - \sigma} ) c^{\dag}_{\vec{i} , \sigma}
  c_{\vec{j} , \sigma} \hat{n}_{\vec{j} , - \sigma} \right]
  \nonumber\\
  &  & + \sum_{<\vec{i} , \vec{j}} \left[
  t_\mathrm{so}^{\vec{i} \vec{j}} \hat{n}_{\vec{i}
  \downarrow} c_{\vec{i} \uparrow}^{\dag} c_{\vec{j} \downarrow}
  \hat{n}_{\vec{j} \uparrow} + t_\mathrm{so}^{\vec{i}
  \vec{j}} ( 1 - \hat{n}_{\vec{i} \downarrow} )
  c_{\vec{i} \uparrow}^{\dag} c_{\vec{j} \downarrow} ( 1 -
  \hat{n}_{\vec{j} \uparrow} ) \right], 
\end{eqnarray}
where $\sigma = \pm 1 / 2$ for spin up (down), respectively. Notice that
$H_\mathrm{SOC}^+$ does not contain Hermitian conjugate, since its hermitian
conjugate processes belong to $H_\mathrm{SOC}^-$. And all the hopping
processes that {\em{conserve}} the number of $\lvert \uparrow \rangle$ are given by
\begin{eqnarray}
  H_\mathrm{SOC}^0 & = & - t_0 \sum_{<\vec{i} ,
  \vec{j}>} \sum_{\sigma} \left[ 2 \sigma ( 1 -
  \hat{n}_{\vec{i} , - \sigma} ) c^{\dag}_{\vec{i}
  , \sigma} c_{\vec{j} , \sigma} ( 1 -
  \hat{n}_{\vec{j} , - \sigma} ) + 2 \sigma
  \hat{n}_{\vec{i} , - \sigma} c^{\dag}_{\vec{i} ,
  \sigma} c_{\vec{j} , \sigma} \hat{n}_{\vec{j} , -
  \sigma} \right] \nonumber\\
  &  & + \sum_{<\vec{i} , \vec{j}>} \left[
  t_\mathrm{so}^{\vec{i} \vec{j}} \hat{n}_{\vec{i}
  \downarrow} c_{\vec{i} \uparrow}^{\dag} c_{\vec{j} \downarrow}
  ( 1 - \hat{n}_{\vec{j} \uparrow} ) + \mathrm{h.c.} \right] . 
\end{eqnarray}
We focus on the low-energy subspace that containing no $\lvert \uparrow \rangle$,
then the projection operator $P_{\uparrow}$ is introduced as
\begin{equation}
  P_{\uparrow} = \prod_i \left[1 - \hat{n}_{i \uparrow} (1 - \hat{n}_{i
  \downarrow})\right],
\end{equation}
to the lowest order, the effective Hamiltonian $H_{\mathrm{eff}}$ is
\begin{eqnarray}
  & H_{\mathrm{eff}} & = P_{\uparrow} (H_{\mathrm{SOC}} +
  H_{\mathrm{Zeeman}} +H_{\mathrm{int}}) P_{\uparrow} \nonumber\\
  &  & = P_{\uparrow} ( H_{\mathrm{SOC}}^0 +
  \tilde{H}_{\mathrm{Zeeman}} ) P_{\uparrow}\nonumber\\
  &  & = - t_0 \sum_{< \vec{i}, \vec{j} >} \left[
  \hat{n}_{\vec{i} \downarrow} \hat{n}_{\vec{j} \downarrow}
  c^{\dag}_{\vec{i} \uparrow} c_{j \uparrow} - ( 1 -
  \hat{n}_{\vec{i} \uparrow} ) c^{\dag}_{\vec{i}
  \downarrow} c_{\vec{j} \downarrow} (1 - \hat{n}_{j \uparrow})
  \right]\nonumber\\
  &  & + \sum_{< \vec{i}, \vec{j} >} \left[
  t_{\mathrm{so}}^{\vec{i} \vec{j}} n_{\vec{i}
  \downarrow} c_{\vec{i} \uparrow}^{\dag} c_{\vec{j} \downarrow}
  ( 1 -\hat{n}_{\vec{j} \uparrow} ) + \mathrm{h.c.} \right] +
  \tilde{H}_{\mathrm{Zeeman}},
\end{eqnarray}
where the compensated Zeeman potential $\tilde{H}_{\mathrm{Zeeman}}$ in
low-energy subspace is
\begin{equation}
  \tilde{H}_{\mathrm{Zeeman}} = \frac{\delta}{2} \sum_{\vec{i}}
  ( \hat{n}_{\vec{i} \uparrow} - \hat{n}_{\vec{i}
  \downarrow} ),
\end{equation}
with $\delta / 2 = 2 m_z - U$.

\section{Technical details and more discussions on the two-body Physics}
We investigate the two-body problem to illustrate the pairing instability of the Fermi surface in the presence of non-Abelian gauge field. We consider a Fermi surface consisting of fully polarized spin-down fermions only, with Fermi energy denoted by $E_F$.

\subsection{The self-consistent equations}
In this subsection, we derive the self-consistent equations for the two-body bound state of fermions. The effective Hamiltonian can be simplified in the case of two-body subspace and the Fermi surface we consider. The spin-up fermions can only be generated by correlated spin-flip processes since the Fermi sea is filled by fully polarized spin-down fermions. Thus the spin-conserved hopping for spin-up particles vanishes, and the term $1-\hat{n}_{\vec{j} \uparrow}$ reduces to $1$ in this regime. Then the real-space Hamiltonian can be rewritten as
\begin{equation}
H_{t}^{0}+\tilde{H}_{\mathrm{Zeeman}}=
    t_{0}\sum_{\langle\vec{i},\vec{j}\rangle}c_{\vec{i}\downarrow}^{\dagger}c_{\vec{j}\downarrow}\\
  +\sum_{\langle\vec{i},\vec{j}\rangle}[t_{\mathrm{so}}^{\vec{i}\vec{j}}\hat{n}_{\vec{i}\downarrow}c_{\vec{i}\uparrow}^{\dagger}c_{\vec{j}\downarrow}+\mathrm{h.c.}]+\frac{\delta}{2}\sum_{\vec{i}}(\hat{n}_{\vec{i}\uparrow}-\hat{n}_{\vec{i}\downarrow}).
\end{equation}
Upon a Fourier transform, the Hamiltonian is casted into
\begin{multline}
H_{t}^{0}+\tilde{H}_{\mathrm{Zeeman}}=\sum_{\vec{k}}\left[2t_{0}(\cos k_{x}+\cos k_{y})n_{\vec{k}\downarrow}+\frac{\delta}{2}(n_{\vec{k}\uparrow}-n_{\vec{k}\downarrow})\right]\\
+\frac{t_{\mathrm{so}}}{N}\sum_{\vec{q},\vec{k},\vec{k'}}\left[\left(-2i\sin k_{x}'\cos\frac{q_{x}}{2}+2\sin k_{y}'\cos\frac{q_{y}}{2}\right)
c_{\frac{\vec{q}}{2}+\vec{k}\uparrow}^{\dagger}c_{\frac{\vec{q}}{2}-\vec{k}\downarrow}^{\dagger}c_{\frac{\vec{q}}{2}-\vec{k'}\downarrow}c_{\frac{\vec{q}}{2}+\vec{k'}\downarrow}+\mathrm{h.c.}\right],
\end{multline}
where $n_{\vec{k},\sigma }=c^\dagger_{\vec{k}\sigma }c_{\vec{k}\sigma} $. 
We then present the details for the trial wave function in the main text. Based on the correlated spin-orbit coupling processes, the two-body trial wave function is a superposition of $s$- and $p$- wave pairs. The projection operator $P_{\uparrow }$ imposes different constraints on the $s$-wave and $p$-wave components of trial wave function in momentum space. The $p$-wave component, consisting of two paired spin-down fermions, remains unchanged by $P_{\uparrow}$. However, for the $s$-wave component, $P_{\uparrow }$ eliminates the momentum-dependence of the up-down pair component of trivial wave function
\begin{equation}
  P_{\uparrow}\left(\sum_{\vec{k}}\psi_s(\vec{k})c_{\frac{\vec{q}}{2}+\vec{k}\uparrow}^\dagger c_{\frac{\vec{q}}{2}-\vec{k}\downarrow}^\dagger\right)=\sum_{\vec{k}}\overline{\psi_s}c_{\frac{\vec{q}}{2}+\vec{k}\uparrow}^\dagger c_{\frac{\vec{q}}{2}-\vec{k}\downarrow}^\dagger,
\end{equation}
with $\overline{\psi_s}=(1/N_0)\sum_{\vec{k}}\psi_s(\vec{k})$ and $\vec{q}$ is the center-of-mass momentum. This is because 
$P_{\uparrow}$ projects out the singly occupied spin-up particles in real space when applied to the up-down pairs, leaving only those pairs with spin-up and spin-down particles occupying the same lattice site.  Thus the $s$-wave component does not have momentum dependence. Combining these ingredients, we write down the trial wave function given in the main text 
\begin{equation}
  \left|\Psi_{\vec{q}}\right\rangle =\left(\sum_{\vec{k}}\psi_{s}c_{\frac{\vec{q}}{2}+\vec{k}\uparrow}^{\dagger}c_{\frac{\vec{q}}{2}-\vec{k}\downarrow}^{\dagger}+\psi_{p}(\vec{k})c_{\frac{\vec{q}}{2}+\vec{k}\downarrow}^{\dagger}c_{\frac{\vec{q}}{2}-\vec{k}\downarrow}^{\dagger}\right)|0\rangle
\end{equation}
Then one can obtain 
\begin{multline}
(H_{t}^{0}+\tilde{H}_{\mathrm{Zeeman}})P_{\uparrow}\left|\Psi_{\vec{q}}\right\rangle=\sum_{\vec{k}}2E_0(\vec{k})\psi_{s}c_{\frac{\vec{q}}{2}+\vec{k}\uparrow}^{\dagger}c_{\frac{\vec{q}}{2}-\vec{k}\downarrow}^{\dagger}|0\rangle
\\+\frac{2t_{\mathrm{so}}}{N}\sum_{\vec{q},\vec{k},\vec{k'}}
\bigg[2\left(-i\sin k_{x}'\cos\frac{q_{x}}{2}+\sin k_{y}'\cos\frac{q_{y}}{2}\right)\psi_p(\vec{k'})c_{\frac{\vec{q}}{2}+\vec{k}\uparrow}^{\dagger}c_{\frac{\vec{q}}{2}-\vec{k}\downarrow}^{\dagger}
\\+\left(i\sin k_{x}'\cos\frac{q_{x}}{2}+\sin k_{y}'\cos\frac{q_{y}}{2}\right)\psi_s c_{\frac{\vec{q}}{2}+\vec{k}\downarrow}^{\dagger}c_{\frac{\vec{q}}{2}-\vec{k}\downarrow}^{\dagger}\bigg]|0\rangle.
\end{multline}
An extra factor of $2$ occurs on the second line due to a double-counting of down-down pair. $E_0(\vec{k})$ is defined as $E_{0}(\vec{k})=2t_{0}\left(\cos k_{x}\text{\ensuremath{\cos (q_x/2)}}+\cos k_{y}\cos (q_{y}/2)\right)-\delta/2$. Hereby, $H_{\mathrm{eff}}P_{\uparrow}\left|\Psi_{0}\right\rangle =E\left|\Psi_{0}\right\rangle $
leads to two equations:
\begin{equation}
-E\psi_{s}=4it_{\mathrm{so}}\frac{1}{N}\sum_{\vec{k'}}\left(\sin k_{x}^{\prime}\cos\frac{q_{x}}{2}+i\sin k_{y}^{\prime}\cos\frac{q_{y}}{2}\right)\psi_{p}(\vec{k'}),\label{Self-Con-1}
\end{equation}
\begin{equation}
\left[2E_0(\vec{k})-E\right]\psi_{p}(\vec{k})\\
=-2it_{\mathrm{so}}\left(\sin k_{x}\cos\frac{q_{x}}{2}-i\sin k_{y}\cos\frac{q_{y}}{2}\right)\psi_{s}.\label{Self-Con-2}
\end{equation}
Then one can obtain the self-consistent equations in the main text 
\begin{equation}
E=-8t_{\mathrm{so}}^{2}\iint_{K}\frac{\mathrm{d}^{2}\vec{k}}{4\pi^{2}}\frac{\sin^{2}k_{x}\cos^{2}\frac{q_{x}}{2}+\sin^{2}k_{y}\cos^{2}\frac{q_{y}}{2}}{2E_{0}(\vec{k})-E},
\end{equation}
and
\begin{equation}
\frac{\psi_{p}(\vec{k})}{\psi_{s}}=-2it_{\mathrm{so}}\frac{\sin k_{x}\cos\frac{q_{x}}{2}-i\sin k_{y}\cos\frac{q_{y}}{2}}{2E_{0}(\vec{k})-E}.
\end{equation}

\subsection{Pair size and effective mass}

In this subsection, we provide the details for the pair size $\ell_C$ and the effective mass $m_B$ for the two-body state. The pair size $\ell_c$ is defined as the root-mean-square average of the distance between the two particles 
\begin{equation}
  \ell_C=\sqrt{\left\langle(\vec{i}-\vec{j})^{2}\right\rangle}.
\end{equation}
Here $\langle \dots \rangle$ is the expectation value of the two-body wave function.
A two-body wave function with zero total momentum written in momentum space can be transformed into real space as
\begin{equation}
\sum_{\vec{k}}f(\vec{k})c_{\vec{k}}^{\dagger}c_{-\vec{k}}^{\dagger}=\frac{1}{N}\sum_{\vec{i},\vec{j}}\tilde{f}(\vec{i}-\vec{j})c_{\vec{i}}^{\dagger}c_{\vec{j}}^{\dagger},
\end{equation}
where $\tilde{f}$ denotes
the Fourier transform of $f$. Then $\ell_C$ can be expressed in terms of the wave function as 
\begin{equation}
\ell_{C}^{2}=\frac{\sum_{\vec{i}}\vec{i}^{2}|\tilde{f}(\vec{i})^{2}|}{\sum_{i}|\tilde{f}(\vec{i})^{2}|}=\frac{\sum_{\vec{k}}|\nabla_{\vec{k}}f(\vec{k})|^{2}}{\sum_{\vec{k}}|f(\vec{k})|^{2}}.
\end{equation}
In our case, we use $f(\vec{k})=\psi_{p}(\vec{k})$ to calculate the pair size, since the s-wave component of our two-body wave function is always on-site.

We use the effective mass $m_B$ to characterize the mobility of the two-body bound state. The dispersion of such bound state is $E(\vec{q})$, where $E$ is the solution of the self-consistent equations. The effective mass $m_B$ is then defined as
\begin{equation}
    m_B=\left(\frac{\partial^{2}E(q)}{\partial q^{2}}\right)^{-1}.
\end{equation}

\section{Mean-field method }
In this section, we provide details for the self-consistent mean-field calculation and additional results supporting the findings in the main text.
\subsection{Projection in the mean-field calculation}
\begin{figure*}[h]
  \includegraphics{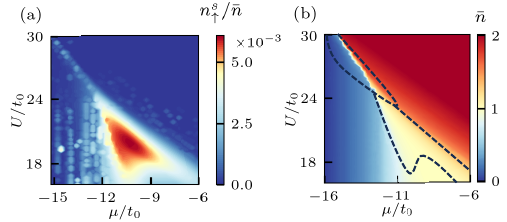}
  \caption{\label{fig:Fig_sm_MF_SingleUp} Additional mean-field result with $t_0=t_{\mathrm{so}}$ and $m_{z}=12t_{0}$. (a) The ratio of singly occupied spin-up states to the average filling number $n_{\uparrow }^{s}/\bar n$. (b) Self-consistent filling number $\bar n$. Dashed area corresponds to the TSF phases. As mentioned in the main text, the upper and lower dashed area correspond to $C=-1$ and $C=+1$ topological superfluid phase, respectively.}
\end{figure*}

\begin{figure*}[h]
  \includegraphics{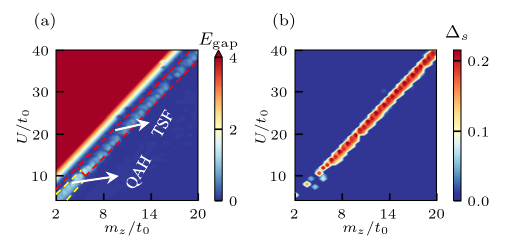}
  \caption{\label{fig:Fig_sm_MF_fixMu_large} Self consistent results keeping $\mu=-mz$. (a) The bulk gap $E_{\mathrm{gap}}$ and (b) $s$-wave pairing potential $\Delta_s$. Red dashed area corresponds to the TSF phases and yellow dashed area corresponds to the QAH phases.}
\end{figure*}

We first show that a large Zeeman splitting associated with a large attractive Hubbard term serves as the projection operator $P_{\uparrow}$ that projects out the singly-occupied spin-up states in the mean-field calculation. The average single-occupation of spin-up particles $n_{\uparrow }^{s}$ can be evaluated by 
\begin{equation}
  n_{\uparrow }^{s}=\langle\hat{n}_{i\uparrow}(1-\hat{n}_{i\downarrow})\rangle	=\langle\hat{n}_{i\uparrow}\rangle-\langle\hat{n}_{i\uparrow}\hat{n}_{i\downarrow}\rangle
 \approx\bar{n}_{\uparrow}-\bar{n}_{\uparrow}\bar{n}_{\downarrow}-|\Delta_{s}|^{2},
\end{equation}
where $\bar n_\uparrow$, $\bar n_\downarrow$ and $\Delta_{s}$ can be obtained from the self-consistent calculation. The effectiveness of the projection is quantified by calculating the ratio of singly-occupied spin-up states to the average particle number $\bar n$. As illustrated in Fig.~\ref{fig:Fig_sm_MF_SingleUp}, the ratio is always less than $1\%$, demonstrating that the singly-occupied spin-up states are indeed projected out by retaining the large Zeeman splitting associated with the large attractive Hubbard term in the mean-field Hamiltonian. Fig.~\ref{fig:Fig_sm_MF_fixMu_large} shows that increasing $m_z$ from moderate regime to large regime can result in the effective projection, since the QAH phase originating from the single-particle spin-flip processes disappears in large $m_z$ regime. Here we perform the self-consistent calculation with $\mu=-m_z$, and increase $m_z$ up to $20 t_0$, which is much larger to the case in the main text. The TSF still persists for even larger $m_z$ as expected from the effectiveness of projection. (A more detailed investigation of $U$-$m_z$ phase diagram can be found in Sec.~\ref{Sec:fixMz}.)

The underlying mechanism is that the large Zeeman term substantially increases the on-site energy of spin-up state $\lvert \uparrow \rangle$, thereby reducing the spin-up states population. The effectiveness of projection is further improved by the large attractive Hubbard term, which lowers the energy of doublon $\lvert \uparrow \downarrow \rangle$, resulting in a larger fraction of the spin-up states originating from doublon. Moreover, our analysis suggests that the effectiveness of the projection can be further improved by increasing the Zeeman splitting $m_z$ associated with Hubbard interaction $U$. 

\subsection{Mean-field Hamiltonian}
In this subsection, we provide the details for the mean-field Hamiltonian. As discussed in the last subsection, we keep the large Zeeman and attractive Hubbard term to effectively project out the singly occupied spin-up states. And we keep only the spin-flip processes to be density dependent, in order to capture the essential pair hopping processes resulted by dynamical gauge field. Combining these ingredients, we consider the following Hamiltonian for the mean-field calculation
\begin{eqnarray}
  & H_{\mathrm{eff}} & = - t_0 \sum_{< \vec{i}, \vec{j} >} (
  c^{\dag}_{\vec{i} , \uparrow} c_{\vec{j} , \uparrow} -
  c^{\dag}_{\vec{i} , \downarrow} c_{\vec{j} , \downarrow}
  ) + \sum_{< \vec{i}, \vec{j} >} \left[
  t_{\mathrm{so}}^{\vec{i} \vec{j}} \hat{n}_{\vec{i}
  \downarrow} c_{\vec{i} \uparrow}^{\dag} c_{\vec{j} \downarrow}
  ( 1 - \hat{n}_{\vec{j} \uparrow} ) + \mathrm{h.c.} \right]
  \nonumber\\
  &  & + m_z \sum_{\vec{i}} ( \hat{n}_{\vec{i} \uparrow} -
  \hat{n}_{\vec{i} \downarrow} ) - U \sum_{\vec{i}}
  \hat{n}_{\vec{i} \uparrow} \hat{n}_{\vec{i} \downarrow} . 
\end{eqnarray}
As discussed in the main text, for the spin-flip term, we consider both the paring channel and the
correlated hopping channel
\begin{eqnarray}
  & \hat{n}_{\vec{i} \downarrow} c^{\dag}_{\vec{i} \uparrow}
  c_{\vec{j} \downarrow} ( 1 - \hat{n}_{\vec{j} \uparrow}
  ) & \approx - \Delta_p^* c_{\vec{i} \uparrow}^{\dag}
  c_{\vec{i} \downarrow}^{\dag} (1 - n_{\uparrow}) - \Delta_s
  c_{\vec{j} \downarrow} c_{\vec{i} \downarrow} (1 - n_{\uparrow}) +
  \Delta_s \Delta_p^* (1 - n_{\uparrow}) \nonumber\\
  &  & + n_{\downarrow} c^{\dag}_{\vec{i} \uparrow} c_{\vec{j}
  \downarrow} (1 - n_{\uparrow}), 
\end{eqnarray}
with $\Delta_s = (1/N_0) \sum_{\vec{i}} \langle c_{\vec{i}
\uparrow}^{\dag} c_{\vec{i} \downarrow}^{\dag} \rangle$, $\Delta_p = (1/N_0)
\sum_{\vec{i}} \langle c^{\dag}_{\vec{i} \downarrow}
c^{\dag}_{\vec{i} + 1  \downarrow} \rangle$ and $\bar n_{\sigma} = (1/N_0)
\sum_{\vec{i}} \langle \hat{n}_{\vec{i} \sigma} \rangle$, which describes the pair hopping processes and renormalized SOC processes resulted by density-dependent gauge field. For the
interaction term, we consider both density channel and pairing channel
\begin{equation}
  \hat{n}_{i \uparrow} \hat{n}_{i \downarrow} \approx \bar n_{\uparrow}
  \hat{n}_{i \downarrow} + \bar n_{\downarrow} \hat{n}_{i \uparrow} -
  \bar n_{\uparrow} \bar n_{\downarrow} + \Delta_s c_{i \downarrow} c_{i \uparrow} +
  \Delta_s^* c^{\dag}_{i \uparrow} c_{i \downarrow}^{\dag} - |
  \Delta_s |^2 .
\end{equation}
Then the Hamiltonian in the Nambu basis $\psi_{\vec{k}} =
(\begin{array}{cccc}
  c_{\vec{k} \uparrow} , & c_{\vec{k} \downarrow} , &
  c^{\dag}_{- \vec{k} \downarrow} , & c^{\dag}_{- \vec{k}
  \uparrow}
\end{array})^{\top}$ can be written as $H_\mathrm{BdG}= (1/2)
\sum_{\vec{k}} \psi^{\dag}_{\vec{k}} \mathcal{H}_{\mathrm{BdG}}
\psi_{\vec{k}}$ up to some constants,
\begin{equation}
  \mathcal{H}_{\mathrm{BdG}} = \left(\begin{array}{cc}
    H_{\mathrm{QAH}}^{\mathrm{MF}} ( \vec{k} ) - \mu & h_1
    \tau_z + 2 h_2 \tau^-\\
    h_1^* \tau_z + 2 h_2^* \tau^+ & - [
    H_{\mathrm{QAH}}^{\mathrm{MF}} ( - \vec{k} ) - \mu ]
  \end{array}\right) ,
\end{equation}
where $\tau^{\pm} = (\tau_x \pm i \tau_y)/2$ and $\tau_{x,y,z}$ are Pauli matrices, and 
\begin{equation}
  H_{\mathrm{QAH}}^{\mathrm{MF}} ( \vec{k} ) =
  \left(\begin{array}{cc}
    h_z ( \vec{k} ) - U n_{\downarrow} & - i h_x (k_x) + h_y
    (k_y)\\
    i h_x (k_x) + h_y (k_y) & - h_z ( \vec{k} ) - U
    n_{\uparrow}
  \end{array}\right),
\end{equation}
with 
\begin{align}
	h_{x}(k_{x}) &=2t_\mathrm{so}\bar n_{\downarrow}(1-\bar n_{\uparrow})\sin(k_x), \nonumber\\
	h_{y}(k_{y}) &=2t_\mathrm{so}\bar n_{\downarrow}(1-\bar n_{\uparrow})\sin(k_y), \nonumber\\
	h_{z}(\vec{k}) &=m_z - 2t_0(\cos(k_x)-\cos(k_y)),\\
	h_1 &=-2t_\mathrm{so}\Delta_p^*(1-\bar n_\uparrow)(1+i)-U\Delta_s ^{*}\nonumber,\\
	h_2(\vec{k}) &=-2 t_{\mathrm{so}}\Delta_s ^{*}(1-\bar n_{\uparrow})(i\sin(k_x)+\sin(k_y)).\nonumber
\end{align}

\subsection{Self-consistent pairing potential}
\begin{figure*}[h]
  \includegraphics{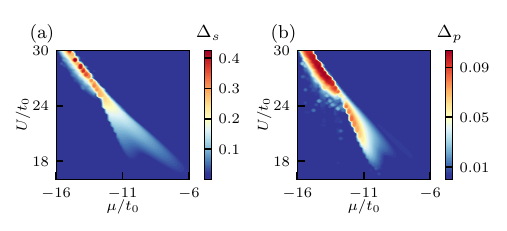}
  \caption{\label{fig:SM_MF_Pair} Self consistent pairing potential (a) $s$-wave pairing potential $\Delta_s$ and (b) $p$-wave pairing potential $\Delta_s$ as a function of chemical potential $\mu$ and interaction $U$. Other parameters are chosen as $t_0=t_{\mathrm{so}}$ and $m_{z}=12t_{0}$.}
\end{figure*}

Fig.~\ref{fig:SM_MF_Pair} shows the self-consistent $s$-wave and $p$-wave order parameters as a function of chemical potential $\mu$ and interaction $U$. We note that although $\Delta_s$ has a larger magnitude than $\Delta_p$, this does not mean the $s$-wave pairing dominates over the $p$-wave pairing; rather the $p$-wave pairing is still significant in the ground state. The non-trivial triplet pairing arises from two sources in this mean-field ansatz, one is from the explicit pair hopping $ \Delta_p^* c_{\vec{i} \uparrow}^{\dag}c_{\vec{i} \downarrow}^{\dag} (1 - n_{\uparrow}) + \Delta_s
c_{\vec{j} \downarrow} c_{\vec{i} \downarrow} (1 - n_{\uparrow})$, and the other comes from the joint action of on-site $s$-wave pairing and the renormalized SOC.

\subsection{Additional mean-field results with $\mu=-m_z$}
\label{Sec:fixMz}

\begin{figure*}[h]
  \includegraphics{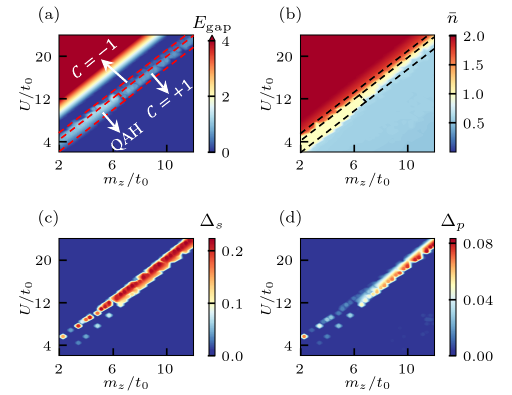}
  \caption{\label{fig:SM_MF_FixMu} Self consistent results keeping $\mu=-mz$ (a) The bulk gap $E_{\mathrm{gap}}$, (b) self-consistent filling number $\bar n$, (c) $s$-wave pairing potential $\Delta_s$ and (d) $p$-wave pairing potential $\Delta_p$ as a function of chemical potential $\mu$ and interaction $mz$. The dashed lines mark the phase boundary of different topological phase. And there is a small range of $C=+1$ we does not mark, but can be seen in the by the presence of nonzero $\Delta_s$ and $\Delta_p$. Other parameters are chosen as $t_0=t_{\mathrm{so}}$. The system has both QAH and TSF phases for moderate $m_z$ and exhibits only TSF phases when $m_z$ is large enough in the relevant fillings. }
\end{figure*}

We show the phase diagram and corresponding pairing potentials as a function of $U$ and $m_z$ in Fig.~\ref{fig:SM_MF_FixMu}, with the chemical potential satisfying $\mu=-m_z$. This can further demonstrates the vanishing quantum anomalous Hall (QAH) phase at large Zeeman splittings $m_z$, complementing the results in the main text with fixed $m_z$. Fig.~\ref{fig:SM_MF_FixMu} shows that around the half-filling $\bar n=1$, the ground state has both insulating QAH phase and topological superfluid (TSF) phase for small and moderate $m_z$. And for large $m_z$ the ground state has a pure TSF phase in the relevant filling. These results are consistent with the results in the main text that (I) the non-Abelian gauge field and associated TSF phases can always be achieved as long as the $m_z$ matches the resonant condition $U\approx 2m_z$. And (II) for the large $m_z$ regime, the system only exhibits TSF phases without other competing insulating phases in the relevant filling, which enhances the reliability of realizing TSF phases in this scheme.

We note that TSF with Chern number $C=-1$ is also abundant because this phase corresponds to a large range of average particle filling $\bar n$. Thus, when $m_z$ is large enough, TSF phases with different Chern numbers $C=\pm 1$ both cover a wide range of fillings. What's more, the phase transition between $C=-1$ and $C=+1$ also happens at $U=2m_z$, which is consistent with the result obtained for the case $m_z=12t_0$ in the main text.

\section{Additional DMRG results}

We perform the density-matrix renormalization group (DMRG) simulation with $U(1)$ symmetry to study the ground state properties of the non-Abelian dynamical gauge field on cylinders for a specified number of particles. We have checked the numerical convergence of our DMRG simulations by comparing the ground state energy $\langle H_{\mathrm{eff}}\rangle$ and the variance of ground state energy $\langle H_{\mathrm{eff}}^2\rangle - \langle H_{\mathrm{eff}}\rangle^2$ for different bond dimensions up to $\chi=1200$.

\begin{figure*}[h]
  \includegraphics{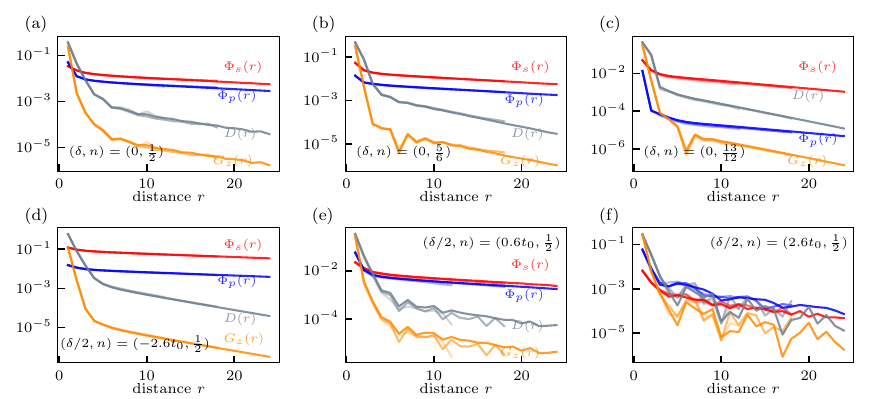}
  \caption{\label{fig:SM_DMRG} Semi-logarithmic plot of
  pair-pair $\Phi_{s/p}(r)$, density-density $D(r)$ and spin-spin correlation
  function $G_z(r)$ with $t_{0}=t_{\mathrm{so}}$ for different
  fillings (a-c) and detunings (d-f).  (a)-(c) correspond to $n=\frac{1}{2},\frac{5}{6},\frac{13}{12}$, respectively, with $\delta=0$. (d)-(f) correspond
  to $\delta=-2.6t_{0},0.6t_{0},2.6t_{0}$, respectively, at $n=\frac{1}{2}$. Negative $\delta $ enhances pair correlations and positive $\delta $ weakens them. In our DMRG simulation, we keep up to 1200 states to obtain accurate results with truncation error $\epsilon\leq5\times10^{-6}$ on cylinders with $L_{y}=4$. Line colors from light to dark represent $L_x=24$, $36$ and $48$.}
\end{figure*}

\begin{figure*}[h]
  \includegraphics{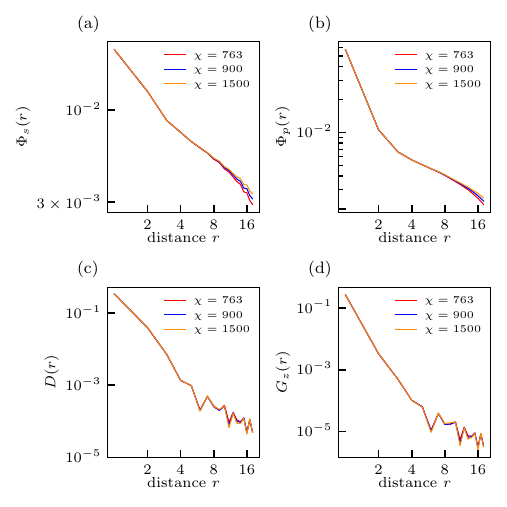}
  \caption{\label{fig:SM_DMRG_bond} Correlation functions for different numbers of bond dimensions $\chi$.  (a) $s$-wave pair-pair correlation function $\Phi_s(r)$, (b) $p$-wave pair-pair correlation function $\Phi_p(r)$, (c) Density-density correlation function $D(r)$ and (d) Spin-Spin correlation function $G_z(r)$ on a $36\times4$ cylinder at detuning $\delta=0.6t_0$ and filling $n=1/2$, with $t_0=t_{\mathrm{so}}$.   All the figures are plotted in the double-logarithmic scales.}
\end{figure*}

In addition to the double-logarithmic plot in the main text, we also present semi-logarithmic plot of correlation functions in Fig.~\ref{fig:SM_DMRG}. The spin-spin correlation function $G_z(r)$ and density-density correlation function $D(r)$ decay faster than $s$-wave and $p$-wave correlation functions, which is consistent with the main text. In the semi-logarithmic plot, $G_z(r)$ and $D(r)$ show very likely a linear decay in the long distance, hinting at the possibility of exponential scalings of $G_z(r)$ and $D(r)$. We note that the same decay behavior shared by $G_z(r)$ and $D(r)$ can be attributed to the presence of spin-orbit coupling of our model, which couples the single-particle charge and spin excitation and make them display the same slope in the long distance.

Fig.~\ref{fig:SM_DMRG_bond} shows the correlation functions for different numbers of bond dimension $\chi$ on a $36\times4$ cylinder at detuning $\delta=0.6t_0$ and filling $n=1/2$ by keeping $\chi=763\sim 1500$ states. It is shown that decay of $\Psi_s(r)$ and $\Phi_p(r)$ becomes slower as the $\chi$ increases, while $D(r)$ and $G_z(r)$ remains unchanged. From these results we expect that the pair-pair correlations should dominate over density-density and spin-spin correlations in the infinite bond dimension.

\end{document}